# NONTRANSITIVE RANKING TO ENHANCE ROUTING DECISION IN MANETS

Md. Amir Khusru Akhtar [1], Arshad Usmani [2], G. Sahoo [3]

[1]Department of Computer Engineering, Cambridge Institute of Technology, Ranchi, Jharkhand, India
[2]Department of Computer Engineering, Cambridge Institute of Technology, Ranchi, Jharkhand, India
[3]Department of Information Technology, Birla Institute of Technology, Mesra, Ranchi, India

**ABSTRACT:**

*An ad hoc network is an infrastructureless network in which nodes perform terminal as well as routing functions. A routing protocol is the only substitute to complete the communications in the absence of an access point. In spite of that mobile nodes or so called routers uses some mechanism for calculating the best route when it has multiple routes for the same destination. On the basis of one or more metrics routes are ranked from best to worst. But, in an ad hoc network many factors can affect this decision, such as the delay, load, route lifetime etc. Thus, measuring and finding routes on the basis of crisp mathematical model for all these attributes is complicated.*

*That's why, the fuzzy approach for best route determination is required for MANET because some of the metrics are fuzzy or vague and the classical ranking of routes and transitivity in the ranking does not hold. The proposed Nontransitive Route Ranking ($NR^2$) model uses the subjective comparison of one route with others and performs nontransitive ranking to rank routes from best to worst. The pairwise comparisons of each route with others give more accurate and fair comparison. The proposed ranking is easier than classical ranking in which metrics have assigned some value and these values are combined to obtain the ranking. Experimental result shows the efficiency of the proposed model.*

## [1] INTRODUCTION

A mobile ad hoc network (MANET) is an infrastructure-less network of mobile nodes or routers connected by wireless links. In this network routing protocol is the only substitute to complete the communications in the absence of an access point. But, due to limited radio range routing in ad hoc network is performed in multi-hop means sending packets through multiple stops instead of one long pathway [1]. Thus, in order to establish a communication, the sender might not communicate directly to the receiver due to the limited radio coverage. So, to route packets devices discover their neighbors to form a network. If the target node is out of range then it is searched by flooding the network with broadcasts that are forwarded by every node. Thus, packets are transmitted through multiple hops to reach the destination. Because, nodes are moving around the network so routing protocols provide the stable connections. Numerous routing protocols have been proposed in the literature such as Ad hoc On-Demand Distance Vector (AODV) [2], Dynamic Source Routing (DSR) [3], Temporally-Ordered Routing Algorithm (TORA) [4], etc., which provide on-the-fly formation to networks. These protocols use some mechanism for calculating the best route when it has multiple routes for the same destination. On the basis of one or more metrics routes are ranked from best to worst [5][6]. But, in an ad hoc network many factors can affect this decision, such as the delay, load, route lifetime etc. Thus, measuring and finding routes on the basis of crisp mathematical model for all these attributes is complicated.

That's why, the fuzzy approach for best route determination is required for MANET because some of the metrics are fuzzy or vague and the classical ranking of routes and transitivity in the ranking does not hold. In this paper we have proposed a Nontransitive Route Ranking ($NR^2$) model which uses the subjective comparison of one route with others and performs nontransitive ranking to rank routes from best to worst. The pairwise comparisons of each route with others give more accurate and fair comparison. The proposed ranking is easier than classical ranking in which metrics have assigned some value and these values are combined to obtain the ranking.

The rest of this paper is organized as follows: Section 2 presents the related work. Section 3 presents the Nontransitive Route Ranking ($NR^2$) model. Section 4 discusses the experiments and results. Finally, Section 5 highlights the conclusion of the paper.

## [2] RELATED WORK

A lot of works have been proposed in literature to enhance routing decision in networks. These works uses various metrics to enhance routing decision in mobile ad hoc network such as the delay, load, route lifetime etc. But, measuring and finding routes on the basis of these all these metrics is complicated because transitivity in the ranking does not hold. That's why we have proposed the nontransitive ranking to enhance routing decision in MANETs. Several works uses fuzzy metrics to enhance routing decision in networks, they are as follows:

The use of fuzzy metric in QoS based OSPF network was proposed by C. K. Tan [7]. In this paper he investigates the relation of two metrics i.e., mean link utilisation and mean link delay, using a fuzzy logic algorithm. He produced a fuzzy metric which is more precise than either of the single metric. But, the limitation of this work is that it focuses only on link utilisation and delay.

A. K. Haboush has proposed a fuzzy optimized metric for adaptive network routing [8]. This paper implements a fuzzy inference rule base to generate the fuzzy cost of each candidate path to be used in routing the incoming calls. He has calculated the fuzzy cost using the crisp values of the different metrics. This work relates three metrics such as mean link bandwidth, queue utilization and mean link delay using a fuzzy logic algorithm to produce an optimized cost of the link. But, the limitation of this work is that it also focuses on limited metrics.

Fuzzy metric approach for route lifetime determination in wireless ad hoc networks was proposed by E. Natsheh et al. [9]. This paper focuses on route lifetime through dynamic measurement instead of static value where route lifetime is called Active Route Timeout (ART) in Ad hoc On-Demand Distance Vector (AODV) routing protocol. They used the fuzzy logic system to study the effect of various parameters on ART. They proposed three design methods for fuzzy ART i.e., fuzzy-SKP, fuzzy-Power and fuzzy-Comb. The proposed fuzzy system optimizes ART efficiently but uses only route lifetime parameter. Hence, the proposed approach has limited application.

An adaptive fuzzy ant-based routing (AFAR) for communication networks was proposed by S. J. Mirabedini et al. [10]. This work takes routing decisions on the basis of fuzzy logic technique with multiple constraints such as path delay and path utilization. Thus, AFAR is capable to take efficient routing decision, but still nontransitivity in routes [11] does not exist. That's why we have proposed the inclusion of nontransitive ranking in this work.

## [3] NONTRANSITIVE ROUTE RANKING (NR$^2$) MODEL

This section presents the Nontransitive Route Ranking (NR$^2$) model which uses the subjective comparison of one route with others and performs nontransitive ranking [11] to rank routes from best to worst. The pairwise comparisons of each route with others give more accurate and fair comparison.

### [3.1] OVERVIEW

The fuzzy approach for best route determination is required for MANET because some of the metrics are fuzzy or vague and the classical ranking of routes and transitivity in the ranking does not hold.

The nontransitivity in ranking is defined as

Let us assume we are ordering routes on the basis of delay, load and route lifetime. When we compare routes $r_1$ to $r_2$, we favor $r_1$; between $r_2$ to $r_3$, we favor $r_2$; but between $r_1$ and $r_3$ we might prefer $r_3$. Thus, transitivity in routes is not maintained. To accommodate this form of nontransitive ranking in routes a special notion of relativity [12] can be used. Let us assume a and b be variables defined on universe M. In order to define a pairwise function $f_b(a)$ when the membership value of a with respect to b is taken and another pairwise function $f_a(b)$ when the membership value of b with respect to a is taken. Now, the relativity function [11][12] known by

$$f(a|b) = \frac{f_b(a)}{\max[f_b(a), f_a(b)]} \qquad (1)$$

This is a measurement of the membership value of choosing a over b. The relativity function f (a | b) can be defined as the preference of variable a over variable b.

The relativity function of one variable means with respect to itself is always identity, i.e.,

$$f(a_i | b_i) = 1 \qquad (2)$$

The matrix of relativity values can be obtained form Eq. 1 has the form f ($a_i$ | $a_j$), where i, j = 1, 2, . . . , n. The obtained matrix should be square matrix and known as comparison matrix (C) of order n.

In order to find the overall ranking, the smallest value in each of the rows of the comparison matrix [11] is defined, i.e.,

$$C'_i = \min f(a_i | M) \quad (3)$$

where i = 1, 2, . . . , n.

The $C'_i$ denotes the membership ranking value for the i[th] variable.

## [3.2] 'C' FUNCTION FOR NONTRANSITIVE ROUTE RANKING (NR$^2$) MODEL

In order to rank the routes from best to worst we have used various metrics such as delay, load, route lifetime, etc. These metrics values are combined and then the subjective measurement of each route with other routes is performed to obtain the pairwise membership function using the Adaptive Fuzzy Ant-based Routing (AFAR) [10]. The pairwise function values are stored in PAIR_MEM_VAL matrix. Then, the relativity values are calculated using Eq. 1 to obtain the comparison matrix (COMP_MATRIX) by relatativity_value_fun() function. After that, the minimum value for each of the rows in comparison matrix is obtained and stored in MIN_VAL_IN_ROW array using minimum_row_fun() function. Finally, the MIN_VAL_IN_ROW array is sorted in descending order using best_to_worst_route_sort_fun() function. The sorted MIN_VAL_IN_ROW array shows the order of the routes from best to worst. The discussed functions are as follows.

```
// C function to Calculate the relativity values to obtain the comparison matrix using Eq. 1

void relatativity_value_fun (double PAIR_MEM_VAL [SIZE][SIZE])
//SIZE denotes number of routes in routing table for the same destination
{
    for (i = 0 ; i <SIZE; i++)
      {
        for (j= 0; j <SIZE; j++)
            {
            if (PAIR_MEM_VAL [i][j]> PAIR_MEM_VAL [j][i])
            large= PAIR_MEM_VAL [i][j];
            else
            large= PAIR_MEM_VAL [j][i];
            COMP_MATRIX [i][j]= PAIR_MEM_VAL [j][i]/large;
             }
        }
}
```

// C function to calculate the minimum value for each of the rows in comparison matrix (COMP_MATRIX) using Eq. 3 and store in (MIN_VAL_IN_ROW) array.

```
void minimum_row_fun (double COMP_MATRIX[SIZE][SIZE])
{
    for(i=0;i<SIZE;i++)
```

```
            {
                  big= COMP_MATRIX[i][0];
                  for(j=1;j<SIZE;j++)
                    {
                          if (COMP_MATRIX[i][j] < big)
                          big= COMP_MATRIX[i][j];
                    }
                  MIN_VALUE_IN_ROW [i]=big;
            }
      }
```

// C function to sort MIN_VAL_IN_ROW array in descending order. The obtained MIN_VAL_IN_ROW array shows the order of the routes from best to worst.

```
void best_to_worst_route_sort_fun(double MIN_VALUE_IN_ROW [SIZE])
  {
  for(i = 1; i < SIZE; i++, b = 0 )
      {
            for(j = 0 ; j < SIZE - i; j++)
            {
             if(MIN_VALUE_IN_ROW [j] < MIN_VALUE_IN_ROW [j+1])
               {
                        temp = MIN_VALUE_IN_ROW [j];
                        MIN_VALUE_IN_ROW [j] = MIN_VALUE_IN_ROW [j + 1];
                        MIN_VALUE_IN_ROW [j + 1] = temp;
                  b++;
               }
            }
          if(b == 0)  break;
      }
  }
```

## [4] EXPERIMENTS AND RESULTS

To obtain results we have performed an experiment to rank the route from best to worst. In order to compare the routes we have used various metrics such as delay, load, route lifetime, etc. These metrics values are combined and then the subjective measurement of each route with other routes is performed to obtain the pairwise membership function using the Adaptive Fuzzy Ant-based Routing (AFAR) [10]. The nontransitive ranking is the requirement for a MANET because of the limited energy and bandwidth constraints. In an ad hoc network many factors can affect this decision, such as the delay, load, route lifetime etc. Thus, measuring and finding routes on the basis of crisp mathematical model for all these attributes is complicated.

Let us consider a node needs to order the routes labeled as $r_1$, $r_2$, $r_3$, $r_4$ and $r_5$ from best to worst. In order to determine which of the route is the most preferred when considered in group rather than in pairs. For this the node first performs the subjective measurement of the appropriateness of each route to determine the pairwise membership functions using the Adaptive Fuzzy Ant-based Routing (AFAR) [10]. The pairwise functions for the routes $r_1$, $r_2$, $r_3$, $r_4$ and $r_5$ are determined as follows:

PAIRWISE FUNCTION VALUES
---------------------------------------------

| 1.00 | 0.50 | 0.30 | 0.20 | 0.70 |
| 0.70 | 1.00 | 0.80 | 0.90 | 0.60 |
| 0.50 | 0.30 | 1.00 | 0.70 | 0.70 |
| 0.40 | 0.10 | 0.60 | 1.00 | 0.80 |
| 0.30 | 0.40 | 0.50 | 0.60 | 1.00 |

Now, the relativity values are calculated using Eq. 1 to obtain the comparison matrix (COMP_MATRIX) by relatativity_value_fun() function. The pairwise functions are determined as follows:

COMPARISON MATRIX
-----------------------------------

| 1.00 | 1.00 | 1.00 | 1.00 | 0.43 |
| 0.71 | 1.00 | 0.37 | 0.11 | 0.67 |
| 0.60 | 1.00 | 1.00 | 0.86 | 0.71 |
| 0.50 | 1.00 | 1.00 | 1.00 | 0.75 |
| 1.00 | 1.00 | 1.00 | 1.00 | 1.00 |

After that, the minimum value for each of the rows in comparison matrix is obtained and stored in MIN_VAL_IN_ROW array using minimum_row_fun() function.

MINIMUM ROW VALUES
---------------------------------------

| 0.43 | 0.11 | 0.60 | 0.50 | 1.00 |

Finally, the MIN_VAL_IN_ROW array is sorted in descending order using best_to_worst_route_sort_fun() function. The sorted MIN_VAL_IN_ROW array shows the order of the routes from best to worst.

ROUTES FROM BEST TO WORST
-------------------------------------------------

  1.00       0.60       0.50       0.43       0.11

Thus, the order of routes from best to worst is $r_5$, $r_3$, $r_4$, $r_1$ and $r_2$. The obtained ordering enhances routing decision in MANETs, because it uses the subjective comparison of one route with others as well as performs the nontransitive ranking to rank routes from best to worst.

## [5] CONCLUSIONS

Existing fuzzy based solutions uses various metrics (such as the delay, load, route lifetime, etc.) to enhance routing decision in mobile ad hoc network. But, measuring and finding routes on the basis of these all these metrics is complicated because transitivity in the ranking does not hold. That's why we have proposed the nontransitive ranking to enhance routing decision in MANETs. In order to rank the routes from best to worst we have used various metrics such as delay, load, route lifetime, etc. These metrics values are combined and then the subjective measurement of each route with other routes is performed to obtain the pairwise membership function using the Adaptive Fuzzy Ant-based Routing (AFAR) [10] algorithm. The pairwise comparisons of each route with others give more accurate and fair comparison. Then, the relativity values are calculated to obtain the comparison matrix which accommodate nontransitive ranking. After that, the minimum value for each of the rows in comparison matrix is obtained which is sorted in descending order to rank routes from best to worst. The proposed ranking is easier than classical ranking. Experimental result shows the efficiency of the proposed model.